\documentclass[preprint,showpacs,preprintnumbers,amsmath,amssymb]{revtex4}
\usepackage{graphicx}% Include figure files
\usepackage{dcolumn}% Align table columns on decimal point
\usepackage{bm}% bold math

\begin{document}
\title{DNA size in confined environments}

\author{Marco Zoli}

\affiliation{School of Science and Technology \\  University of Camerino, I-62032 Camerino, Italy \\ marco.zoli@unicam.it}

\date{\today}

\begin{abstract}
For short DNA molecules in crowded environments, we evaluate macroscopic parameters such as the average end-to-end distance and the twist conformation by tuning the strength of the site specific confinement driven by the crowders. The ds-DNA is modeled by a mesoscopic Hamiltonian which accounts for the three dimensional helical structure and incorporates fluctuational effects at the level of the base pair. The computational method assumes that the base pair fluctuations are temperature dependent trajectories whose amplitudes can be spatially modulated according to the crowders distribution. We show that the molecule elongation, as measured by the end-to-end distance varies non-monotonically with the strength of the confinement. Furthermore it is found that, if the crowders mostly confine the DNA mid-chain, the helix over-twists and its end-to-end distance grows in the strong confinement regime. Instead, if the crowders mostly pin one chain end, the helix untwists while the molecule stretches for large confinement strengths. Thus, our results put forward a peculiar relation between stretching and twisting which significantly depends on the crowders profile. The method could be applied to disegn specific DNA shapes by controlling the environment which constrains the molecule. 
\end{abstract}


\maketitle

\section*{I. Introduction}

Significant advances in nano-channels fabrication make nowadays possible to study the properties of long DNA chains in confined environments under various physical regimes which can be sampled basically by tuning the channel size with respect to the molecule persistence length \cite{austin12,doyle16,odi08,dekker08}. In fluidic chips, DNA sequences are first uncoiled and then driven through nano-channel arrays which uniformly stretch the molecules thus permitting genome mapping and high resolution detection of structural variations possibly involved in genomic rearrangements and disorders \cite{kwok12}, also identified through paired-end-tags sequencing \cite{eichler08,ruan09,campbell11}.  Moreover, nano-channel confinement combined with fluorescence microscopy has been applied \textit{i)} to construct accurate distributions of distances between fluorescent labels by aligning tens of thousands of barcoded DNAs to a reference genome \cite{dorf15}; \textit{ii)} to 
perform DNA denaturation mapping after heating the molecule at temperatures in which single (ss) and double stranded (ds) regions coexist due to different sequence compositions. Then staining the molecule with a fluorescent dye that binds only to ds regions one obtains an optical barcode  corresponding to the specific denaturation pattern of the molecule to be compared with the theoretical melting profiles \cite{reisner10}. 

The DNA denaturation properties are largely affected in confined conditions as those which occur \textit{in vivo} in the cellular crowded environments whereby organelles and macro-molecules reduce the free volume available to base pair fluctuations thus suppressing the melting entropy \cite{nakano14}. Molecular dynamics simulations for DNA oligomers of various length and sequence have shown that macro-molecular crowding stabilizes the base pair hydrogen bonds against thermal disruption \cite{liang14,bozorg15}, an effect which can substantially increase the efficiency of polymerase-chain-reactions by favoring primer-template binding in the annealing and extension phases  \cite{ragh09}. It has also been observed that the DNA thermal stability depends on the crowder structure, concentration and molecular weight as well as  on the sequence length with longer oligomers displaying  higher melting temperatures $T_m$ \cite{goobes03}. To account for these properties, molecular thermodynamic models have been developed  which treat DNA as a freely jointed chain of poly-ions immersed in salted and crowded aqueous solution \cite{jiang07,jiang12}. Each pair of nucleotides in the ds-DNA (each nucleotide in the ss-DNA) is represented by coarse-grained charged particles interacting with crowders, water molecules and salt ions, via Lennard-Jones (LJ) and Coulomb potentials.  It has been shown that $T_m$ increases with the crowders size and concentration whereas, tuning the depth of the LJ potential well, $T_m$ may decrease by enhancing the concentration of specific crowders (as observed experimentally with poly-ethylene glycol \cite{nakano04}) if the interactions of the latter with two separate single strands are energetically favored with respect to those with ds-DNA.  
Also statistical methods based on a simple ladder Hamiltonian model have been applied to derive $T_m$ and the site dependent base pair opening probabilities, for a short ds-DNA molecule, in crowded environments \cite{singh17}: the effect of the crowders size has been simulated by tuning the base pair dissociation energy for the hydrogen bonds described by a Morse potential and, for a given size,  $T_m$ has been found to scale linearly with the crowder density.

More generally, macro-molecular crowding affects the diffusion of molecules in living cells and the dynamics of DNA looping \cite{schleif92}. This, in turn, may reduce the search time for site specific DNA-binding proteins \cite{elf09,dekker12,marko15} and control the speed of gene activation or repression as shown e.g., in synthetic cellular systems made of phage T7 molecular components \cite{leduc13}.  Also the DNA looping kinetics depends on the crowder size with larger crowders favoring polymer confinement and packaging which leads to enhanced looping probabilities, although the latter may significantly depend on the chain length and stiffness \cite{cherst15b}. 

While an extensive body of experimental research is investigating the relation among macro-molecular crowding, DNA dynamics and its biological functioning, much less  theoretical work has been done so far on nucleic acids in crowded conditions also in view of the difficulty of simulating the effects of 
intra-cellular environment on the thermodynamics and kinetics of nucleic acids. 
Here we contribute to this field focusing on the interplay between crowding and DNA structure, namely analyzing how the real space confinement of a DNA molecule due to the presence of crowders may concur to shape the helical conformation and the overall size of the molecule itself. Our study is based on a statistical mechanical method which has been widely discussed in previous researches on some fundamental indicators of DNA flexibility at short length scale such as the cyclization probabilities, the persistence lengths \cite{io16b,io18b} and also the single molecule response to stretching perturbations induced by external loads \cite{io16a,io18a}. 

The method computes the equilibrium conformation for a molecule described by a three dimensional mesoscopic model which contains the intertwined effects of the radial and angular degrees of freedom. In this regard, our approach provides a representation of the DNA structure more realistic than the above mentioned simple ladder models. Moreover, as the Hamiltonian contains fluctuational effects at the level of the single base pairs, we can tune the effects of the crowders operating a site dependent confinement of the base pair fluctuational amplitudes. 
Importantly, the model explicitly incorporates large twisting and bending fluctuations between adjacent base pairs along the molecule stack which are crucial in the calculation of the DNA flexibility properties \cite{maiti15,menon13,york05,tan17}. These features provide a significant improvement, as for the description of the structure of a helical polymer, also over the simple bead-spring chain models  \cite{chen13}  usually adopted to simulate the DNA dynamics in crowded environments \cite{cherst15a}. 

The paper is organized as follows: in Section II, we review the geometrical representation of the helical molecule and the key features of the mesoscopic Hamiltonian model. Section III contains the general features of the statistical method based on the finite temperature path integration. Section IV presents the crowders distribution and the details of the computational method while the results are discussed in Section V. Some final remarks are made in Section VI.

\section*{II. Model }

\subsection*{A. Helix }

\begin{figure}
\includegraphics[height=8.0cm,width=8.0cm,angle=-90]{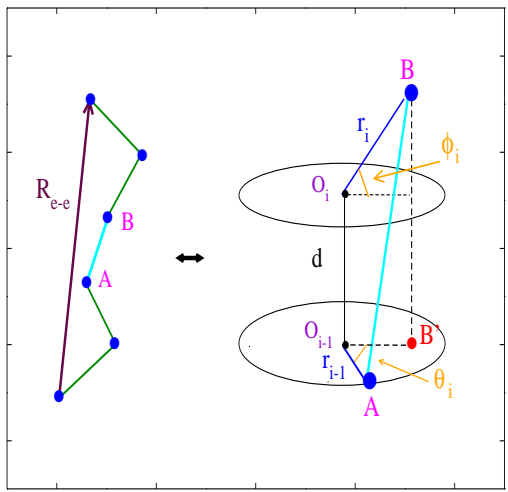}
\caption{\label{fig:1}(Color online)  
Schematic of the three dimensional model for a linear chain with $N$ point-like base pairs. $R_{e-e}$ is the end-to-end distance. The segment $\overline{AB}$, magnified in the right panel, represents the separation between the tips of the radial displacements $r_{i}$, $r_{i-1}$ of adjacent base pairs along the chain backbone. 
In turn, the $r_{i}$'s are the inter-strand fluctuational amplitudes between the $i-th$ base pair' mates. The $r_{i}$'s are measured with respect to the $O_i$'s which lye along the helix mid-axis. Adjacent $O_i$'s are separated by a constant length $d$. 
$\theta_i$ and $\phi_{i}$ are the local twist and bending angles between adjacent base pairs. 
By suppressing the bending fluctuations, our model would map onto a fixed-plane helical model as described by the ovals in the right panel.
}
\end{figure}

Our study is based on a three dimensional helical model for the open end chain made of $N$ base pairs as depicted in Fig.~\ref{fig:1}.  For each base pair, the distance $\, r_{i}$ between the pair mates is the variable accounting for the radial fluctuations with respect to the bare helix diameter $R_0$. The variable range is taken consistently with the physical properties of the DNA molecule. For instance, while radial fluctuations may also lead to local contractions of the inter-strand separation i.e., $\, |r_{i}| < R_0$,  such separation encounters a lower cutoff as complementary strands cannot come too close to each other due to the repulsion between negatively charged phosphate groups. On the other hand, radial displacements much larger than $R_0$ may cause local pair breaking and formation of fluctuational bubbles along the double helix \cite{io14a}. Crowded environments have generally the effect to reduce the amplitude of the base pair fluctuations.
These constraints are implemented in the model potential and in the computational method as discussed below.
Upon suppression of the stretching fluctuations, the blue dots in Fig.~\ref{fig:1} would map onto the $O_i$'s which are arranged at a fixed rise distance $d$ along the helix mid-axis. Accordingly the model would reduce to a freely jointed chain with $N-1$ bonds. In the following, the bare helix parameters are set to: $\,R_0=\,20 \, \AA$ and $d=\,3.4 \, \AA$.

As the $N$ point-like objects (blue dots) are linked by covalent bonds due to effective stacking forces stabilizing the helix, each point precisely describes the monomer unit of DNA  whereas the nucleotides internal degrees of freedom are not considered here.
Furthermore, our representation contains two other variables that is, the twist angle $\theta_i$ and the bending angle $\phi_{i}$ between adjacent base pair fluctuations,  $r_{i}$ and $r_{i-1}$.  It is finally noticed that, by suppressing the bending fluctuations, the model would transform into a fixed-plane helical representation as previously discussed in ref.\cite{io11}.

\subsection*{B. Hamiltonian }

The geometrical model in Fig.~\ref{fig:1} is quantitatively treated by a mesoscopic Hamiltonian which accounts for the main forces stabilizing the double stranded helix i.e., \textit{1)} the hydrogen bonds between base pair mates, described by a one particle potential $V_{1}[r_i]$ and \textit{2)} the intra-strand stacking forces between adjacent base pairs along the molecule stack, described by a two particles potential $V_{2}[ r_i, r_{i-1}, \phi_i, \theta_i]$.  Explicitly,  for a helix made of $N$ base pairs of reduced mass $\mu$, the Hamiltonian reads:

\begin{eqnarray}
& &H =\, H_a[r_1] + \sum_{i=2}^{N} H_b[r_i, r_{i-1}, \phi_i, \theta_i] \, , \nonumber
\\
& &H_a[r_1] =\, \frac{\mu}{2} \dot{r}_1^2 + V_{1}[r_1] \, , \nonumber
\\
& &H_b[r_i, r_{i-1}, \phi_i, \theta_i]= \,  \frac{\mu}{2} \dot{r}_i^2 + V_{1}[r_i] + V_{2}[ r_i, r_{i-1}, \phi_i, \theta_i]  \, . 
\label{eq:01}
\end{eqnarray}

The term $H_a[r_1]$ is separated from the sum as the first base pair lacks the preceding neighbour along the molecule stack. More specifically, $V_{1}$ and $V_{2}$ are written as:

\begin{eqnarray}
& &V_{1}[r_i]=\, V_{M}[r_i] + V_{Sol}[r_i] \, , \nonumber
\\
& &V_{M}[r_i]=\, D_i \bigl[\exp(-b_i (|r_i| - R_0)) - 1 \bigr]^2  \, , \nonumber
\\
& &V_{Sol}[r_i]=\, - D_i f_s \bigl(\tanh((|r_i| - R_0)/ l_s) - 1 \bigr) \, , \nonumber
\\
& &V_{2}[ r_i, r_{i-1}, \phi_i, \theta_i]=\, K_S \cdot \bigl(1 + G_{i, i-1}\bigr) \cdot \overline{d_{i,i-1}}^2  \, , \nonumber
\\
& &G_{i, i-1}= \, \rho_{i, i-1}\exp\bigl[-\alpha_{i, i-1}(|r_i| + |r_{i-1}| - 2R_0)\bigr]  \, . 
\label{eq:02}
\end{eqnarray}

The one particle potential is the sum of a  $V_{M}[r_i]$ and a solvent potential $V_{Sol}[r_i]$.  $V_{M}$ is the effective term for the hydrogen bonds measuring the pair interaction energy from the minimum value (defined as the state with no fluctuations, i.e., $|r_i| = R_0$) up to the pair dissociation energy $D_i$. Due to its repulsive core, $V_{M}$ also accounts for the inter-strands interactions driven by the phosphate groups and, through its parameters, controls the range of the radial fluctuations included in the computation. In fact, the program discards those short radial fluctuations such that $V_{M}[r_i] >  D_i $ \cite{n1} yielding a large electrostatic repulsion hence, a small statistical weight in the partition function.  
On the other hand,  if the base pair fluctuations are large with respect to the bare helix diameter i.e., $|r_i| \gg  R_0$, then the pair mates would sample the flat part of the Morse potential and, in principle, they could go far apart with no further energy cost. This situation however does not account for those recombination
events which instead may take place in solutions and calls for corrections to the physical picture provided by $V_{M}$. 
Furthermore, when a base gets out of the stack, it may form a hydrogen bond with the solvent and, in order to re-close, the base encounters an entropic barrier, not described by $V_M$. For these reasons, the one particle term also contains the term $V_{Sol}$ which increases by  $D_i f_s$ the energy threshold for base pair breaking and defines the barrier, whose width is tuned by $l_s$, which controls the strand recombination occurring in solution.
Although the salinity effects are not studied here, $f_s$ can be related to the salt concentration in the solvent {which is known to affect the overall DNA conformation \cite{livolant16}}.  This link is empirically established by noticing that the DNA melting temperatures scale linearly with the pair dissociation energy and scale logarithmically with the sodium concentration \cite{blake98,lucia98} hence, the height of the potential barrier is also assumed to vary logarithmically with the salt concentration \cite{io11}. 

We take hereafter potential parameters used in recent works on homogeneous fragments i.e., $D_i=\,60 meV$, $b_i= 5 \AA^{-1}$,  $f_s=\,0.1$, $l_s=\,0.5 \AA$ \cite{io18c} and consistent with those obtained by fitting thermodynamic and elastic data  \cite{campa98,krueg06,singh11}.

The two particles potential $V_{2}$ depends on the angular variables through the distance $\overline{d_{i,i-1}}$, corresponding to the segment $\overline{AB}$ in Fig.~\ref{fig:1}, which can be straightforwardly derived in analytic form. Such dependence has some fundamental implications for our model: 

\textit{a)} $V_{2}$ in  Eq.~(\ref{eq:02}) does not vanish in the presence of the translational mode, i.e., when all $r_i$'s are equal. As a consequence, the partition function does not diverge for large $r_i$'s. In contrast, for Hamiltonian models representing ds-DNA as a ladder \cite{pey04}, the absence of angular variables lets $V_{2}$ vanish for the zero mode whereas the on-site potential $V_{1}$ is bound and lacks of translational invariance. Thus, the ensuing divergence of the partition function cannot be removed as usually done for translationally invariant systems and should rather be tackled by a truncation of the phase space  which always carries some arbitrariness as for the choice of the integration cutoffs \cite{pey95,zhang97,kalos09}. 

\textit{b)} Thermal fluctuations may flip a base out of the stack thus reducing the overlap between $\pi $ electrons on adjacent sites. If the separation between adjacent bases becomes very large, in a ladder model, $V_{2}$  diverges \cite{joy05}. Physically however the stacking interactions are finite due to the stiffness of the covalent bonds linking nucleotides in the sugar-phosphate backbone. Our model accounts for the finiteness of the stacking energy through the angular variables in $\overline{d_{i,i-1}}$ which render $V_{2}$ stable against thermal disruptions \cite{io12}. 

\textit{c) }In terms of the twist variable, the code generates a large number of possible conformations, each identified by a value for the average helical repeat, i.e., the number of base pairs per helix turn \cite{io17}. The gist of the method lies in the determination of the equilibrium twist conformation and in the computation of the helical properties as a function of the twist conformation as shown below.

$V_{2}$ is expressed in terms of a elastic force constant $K_S$ and non-linear parameters $\rho_{i, i-1}, \, \alpha_{i, i-1}$ which have been first introduced to account for the cooperativity effects in the formation of thermal fluctuational bubbles and for the sharp DNA denaturation transition \cite{pey93,ares05,rapti06,kalos11}. The range of parameters values has been widely discussed in relation to the computation of thermodynamic and cyclization properties of short sequences \cite{io16b}.  As the focus is here on the effects of the molecule confinement, we can neglect the sequence specificities and assume the homogeneous set of stacking parameters also taken in ref.\cite{io18c}
: \, $K_S=\,10 mev \AA^{-2}$, $\rho_{i} \equiv \, \rho_{i, i-1} =\,1$, $\alpha_{i}\equiv \, \alpha_{i, i-1} =\, 2 \AA^{-1}$. 

In general, both the one particle and two particles potentials can be used to model also heterogeneous sequences through appropriate choices of the parameters as shown e.g., in mesoscopic Hamiltonian studies of end fraying effects in short duplexes  \cite{weber15}, helix untwisting of circular DNA \cite{io13} and thermal stability of nucleic acids hybrids \cite{weber19}.

\section*{III. Method}

The model in Eqs.~(\ref{eq:01}),~(\ref{eq:02}) is studied by a path integral method \cite{fehi} which basically treats the base pair fluctuations as imaginary time dependent paths
$r_i(\tau)$ with $\tau \in [0, \beta]$ and  $\beta=\,(k_B T)^{-1}$. $k_B$ is the Boltzmann constant and $T$ is the temperature. As the method has been described in several papers, we skip here the details and refer e.g., to Refs. \cite{io16b,io18c} for a broad analysis.
For the current purposes it should be mentioned that, by virtue of the space-time mapping technique,  the partition function $Z_N$ for the system in  Fig.~\ref{fig:1} is given by an integral over closed paths, $r_i(0)=\,r_i(\beta )$. As a consequence, the paths $r_i(\tau)$ can be expanded in a Fourier series, \, $r_i(\tau)=\, (r_0)_i + \sum_{m=1}^{\infty}\Bigl[(a_m)_i \cos( \frac{2 m \pi}{\beta} \tau ) + (b_m)_i \sin(\frac{2 m \pi}{\beta} \tau ) \Bigr]$, \, whose coefficients yield, for any base pair, a set of possible fluctuations statistically contributing to $Z_N$. The code includes in the computation only those combinations of Fourier coefficients yielding radial fluctuations which are physically consistent with the constraints imposed by the model potential as explained in Section II.  

Carrying out the integral over the Fourier coefficients, an increasing number of trajectories is added to $Z_N$ until the latter numerically converges.  As the model contains angular degrees of freedom, the convergence is checked also against the bending and twisting fluctuations. This procedure ultimately determines the state of thermodynamic equilibrium for the system.

Explicitly,  for the Hamiltonian in Eq.~(\ref{eq:01}),  $Z_N$ reads:

\begin{eqnarray}
& &Z_N=\, \oint Dr_{1} \exp \bigl[- A_a[r_1] \bigr]   \prod_{i=2}^{N}  \int_{- \phi_{M} }^{\phi_{M} } d \phi_i \int_{- \theta_{M} }^{\theta _{M} } d \theta_{i} \oint Dr_{i}  \exp \bigl[- A_b [r_i, r_{i-1}, \phi_i, \theta_i] \bigr] \, , \nonumber
\\
& &A_a[r_1]= \,  \int_{0}^{\beta} d\tau H_a[r_1(\tau)] \, , \nonumber
\\
& &A_b[r_i, r_{i-1}, \phi_i, \theta_i]= \,  \int_{0}^{\beta} d\tau H_b[r_i(\tau), r_{i-1}(\tau), \phi_i, \theta_i] \, ,
\label{eq:03}
\end{eqnarray}

The bending and twisting integration cutoffs are taken as $\phi_{M}=\,\pi /2$ and $\theta_{M}=\,\pi /4$, respectively. Thus, Eq.~(\ref{eq:03}) includes large fluctuational angles allowing local distortions and formation of kinks which preserve the base pairing but reduce the bending energy \cite{crick75,kim14}. Kinks are expected to increase the cyclization efficiency, markedly in short chains, possibly due to the molecule asymmetric structure \cite{sung15,ejte15}. 

Consistently with the  path Fourier expansion mentioned  above,  $\oint {D}r_i$ in Eq.~(\ref{eq:03}), defines the integration measure over the Fourier coefficients:

\begin{eqnarray}
& &\oint {D}r_{i} \equiv {\frac{1}{\sqrt{2}\lambda_{cl}}} \int_{-\Lambda_{T}^{0}}^{\Lambda_{T}^{0}} d(r_0)_i \prod_{m=1}^{\infty}\Bigl( \frac{m \pi}{\lambda_{cl}} \Bigr)^2 \int_{-\Lambda_{T}}^{\Lambda_{T}} d(a_m)_i \int_{-\Lambda_{T}}^{\Lambda_{T}} d(b_m)_i \, , \, 
\label{eq:04}
\end{eqnarray}

with  $\lambda_{cl}$ being the classical thermal wavelength. While in principle the cutoffs \, $\Lambda_{T}^{0}$ and  $\Lambda_{T}$ should be infinitely large, for computational purposes they are set to finite values by noticing that, as a key property of the finite temperature path integration,  $\oint {D}r_i$  normalizes the kinetic action \cite{io04}: 

\begin{eqnarray}
\oint {D}r_i \exp\Bigl[- \int_0^\beta d\tau {\mu \over 2}\dot{r}_i(\tau)^2  \Bigr] = \,1 \, .
\label{eq:05} \,
\end{eqnarray}

Then, from Eqs.~(\ref{eq:04}),~(\ref{eq:05}) and using the path Fourier expansion, one derives the explicit integration cutoffs in the space of the Fourier coefficients:

\begin{eqnarray}
& &{\Lambda^0_T}=\,\lambda_{cl}/\sqrt{2} \, \nonumber
\\
& &\Lambda_T =\,{{U \lambda_{cl}}  \over {m \pi^{3/2}}},
\,
\label{eq:06}
\end{eqnarray}

with $U$ being a dimensionless parameter. Setting $U=\,12$, one fulfills Eq.~(\ref{eq:05}) and also includes large amplitude base pair fluctuations in the computation of Eq.~(\ref{eq:03}).

In Section IV,  a method is devised to quantify the effects of crowders within the computational scheme described in Eqs.~(\ref{eq:03}),~(\ref{eq:04}).

\section*{IV. Crowding}

\begin{figure}
\includegraphics[height=8.0cm,width=8.0cm,angle=-90]{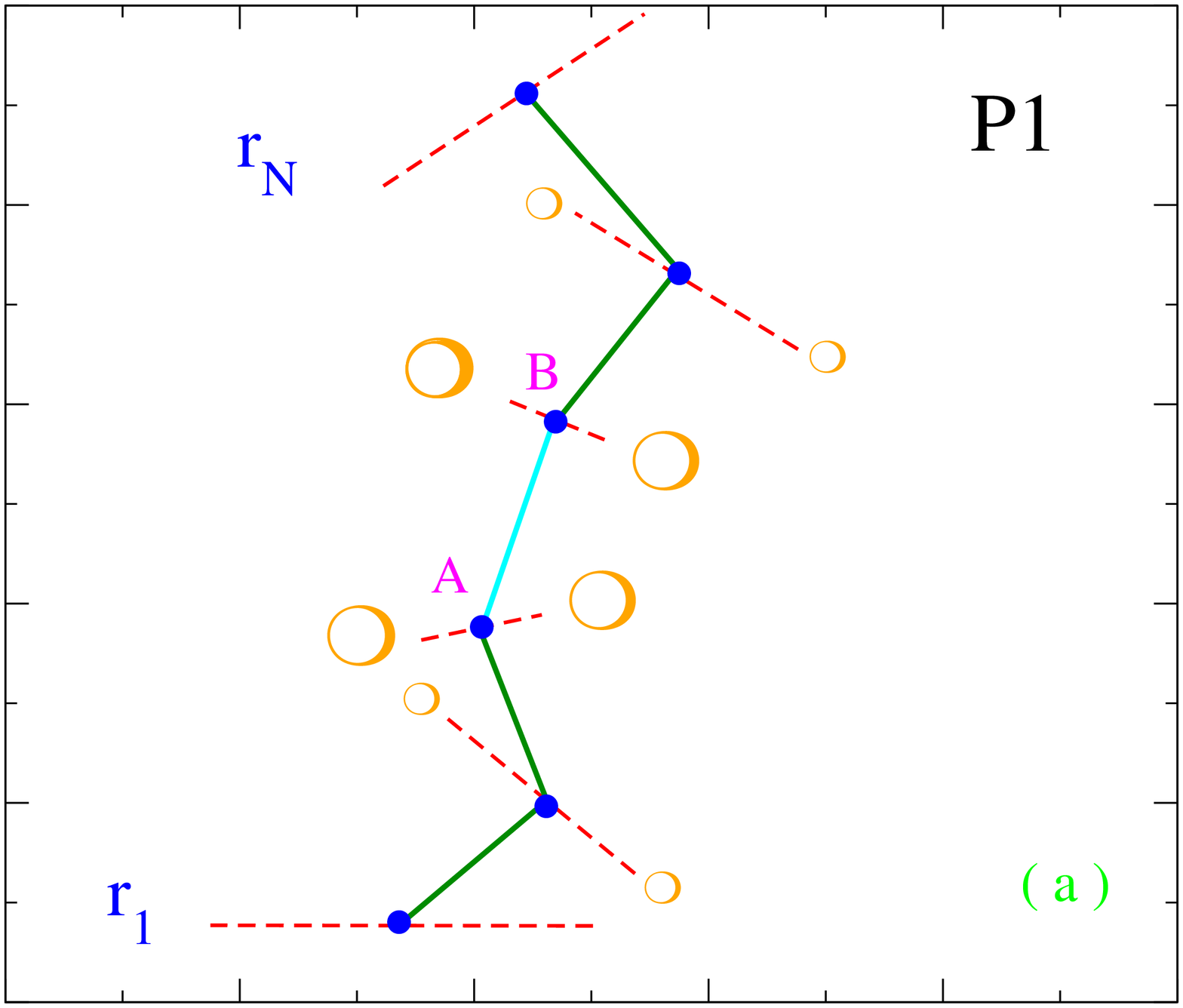}
\includegraphics[height=8.0cm,width=8.0cm,angle=-90]{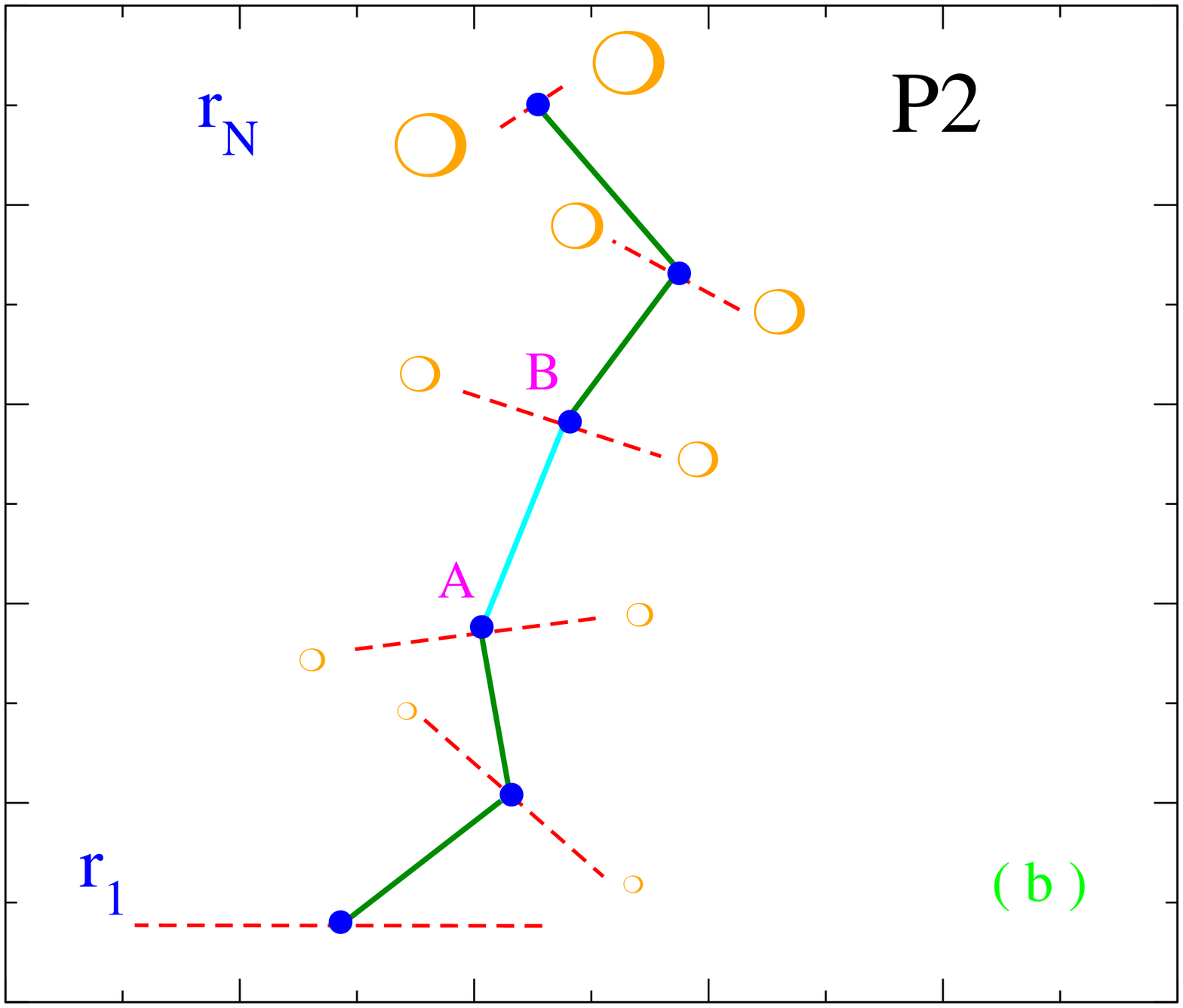}
\caption{\label{fig:2}(Color online)  
Crowders, modeled as spherical objects of variable size, confine the DNA molecule reducing the amplitude of the base pair fluctuations (red dashed lines). Larger size crowders more effectively contract the site dependent fluctuations. Two crowding profiles are considered: (a)\textbf{ P1}. The confinement is maximum at the mid-chain and linearly decreases moving from the middle to the chain ends. For the terminal base pairs, the confinement vanishes.
(b) \textbf{P2}. The confinement is maximum for the last base pair in the chain and linearly decreases going towards the opposite end. The first base pair feels no confinement due to the crowders distribution.}
\end{figure}

As macro-molecular crowding is expected to restrict the free space available to the DNA molecule, we study the helix in a confined environment assuming two distinct distributions of crowders as depicted in Fig.~\ref{fig:2}. In the first profile, left panel, the largest objects exerting the maximum confinement are found around the mid-chain while the size of the crowders around the helix gradually shrinks and finally vanishes as one approaches the chain ends. In the second profile, right panel, the largest crowders are arranged around the chain end so that the base pair fluctuation $r_{N}$ feels the maximum confinement. Moving along the chain, the crowders size shrinks and eventually vanishes around the first base pair which, accordingly, feels no confinement. 
Note that the spherical objects in Fig.~\ref{fig:2} do not represent macro-molecules neighboring single base pairs of the DNA chain but rather simulate a distribution of crowders exerting a variable degree of confinement along the chain.
While these profiles are merely speculative and may not have a counterpart in real systems, they offer a useful model to study the molecule response to site specific reductions of the base pair fluctuations.

Then, we proceed by replacing $U$ by a site dependent cutoff, $\, U \rightarrow U - C(i)$, where  $C(i)$ is the function simulating the effects of the crowders distribution. The profile in Fig.~\ref{fig:2}(a) is defined by:

\begin{eqnarray}
C_{P1}(i)  =\,
\left\{\begin{matrix}
& &   2 \gamma U   \Bigl(\frac{i-1}{N-1} \Bigr)    \hskip 3cm           1 \leq i \leq  (N+1)/2  \\ 
& &    2 \gamma U  \Bigl(\frac{N-i}{N-1}  \Bigr)    \hskip 3cm    (N+1)/2 \leq i \leq N
\end{matrix} \right. \, 
\label{eq:07}
\end{eqnarray}

while the linear crowders distribution in Fig.~\ref{fig:2}(b) is modeled by:

\begin{eqnarray}
C_{P2}(i)  =\, \frac{\gamma U }{N-1} \bigl(i - 1 \bigr)  \hskip 3cm     1 \leq i \leq N   \, , 
\label{eq:08}
\end{eqnarray}

where $\gamma$ is a tunable parameter. For both profiles, we set $\Gamma  \equiv \, U (1 - \gamma)$ as the cutoff measuring the maximum reduction of the base pair radial fluctuations. Only, for the case P1, $\Gamma $ is the mid-chain cutoff while, for the case P2,  $\Gamma $ defines the cutoff at the chain end.

Eqs.~(\ref{eq:07}),~(\ref{eq:08}) are implemented in the program which computes the DNA size and shape in thermal equilibrium with the crowded environment.
Specifically, the ensemble averages for the molecule physical properties are carried out via Eqs.~(\ref{eq:03}),~(\ref{eq:04}),~(\ref{eq:06}) using the cutoffs in Eqs.~(\ref{eq:07}),~(\ref{eq:08}). Thus, for each $\gamma$, the program searches for the state of thermodynamic equilibrium as outlined above.

As \textit{a priori} the twist conformation of the short fragment is unknown, we sample a broad range of J-values for the number of base pairs per helix turn,  around the experimental $\,h^{exp}=\,10.4$ value, albeit estimated and generally accepted for kilo-base long DNA \cite{wang79}. 

For the $j-th$ value in such range (j=\,1,...,J), we apply a iterative procedure which  accounts for the helical shape of the molecule and also yields the ensemble averaged twist angles $<\theta_{i}>$ for all base pairs in the chain. Explicitly
the $i-th$ twist angle in Fig.~\ref{fig:1} is written as, \, $\theta_i =\, <\theta_{i - 1}>  + 2\pi / h_j + \theta_{i}^{fl}$, \, where $<\theta_{i - 1}>$ is the ensemble averaged twist for the preceding base pair along the stack (\, $< \theta_1 >\equiv \,0$ )
and  $\theta_{i}^{fl}$ is the twist fluctuation variable. This means that the twist integration in Eq.~(\ref{eq:03}) is performed over $\theta_{i}^{fl}$. Then the output of one iterative step ($<\theta_{i - 1}>$) is the input for the next. Once $< \theta_N >$ is obtained, one gets the $j-th$ helical repeat averaged over the statistical ensemble of Eq.~(\ref{eq:03}):

\begin{eqnarray}
< h >_{j}=\,\frac{2\pi N}{< \theta_N >} \, . 
\label{eq:09}
\end{eqnarray}

As, in turn, the procedure is repeated for any initial $h_j$ value, we then derive a set of J- average twist conformations. Among the latter, the value ($< h >_{j^{*}}$)  that corresponds to the state of thermodynamic equilibrium is finally selected by minimizing the free energy $\, F=\, -\beta ^{-1} \ln Z_N$.
For any twist conformation defined by Eq.~(\ref{eq:09}), the computational method can also deliver the macroscopic average parameters providing a measure for the global DNA size. In particular, we focus hereafter on the average end-to-end distance calculated as:

\begin{eqnarray} 
< R_{e-e} > =\, \biggl < \biggl| \sum_{i=2}^{N}  \overline{d_{i,i-1}} \biggr| \biggr >   \,.
\label{eq:10}
\end{eqnarray}

The outlined strategy  holds for one $\gamma$ (or $\Gamma$) that is, for one crowders distribution. Tuning $\Gamma$, we further monitor the changes in the molecule shape induced by the confining environment.

For one $\Gamma$, the execution time depends on  \textit{i)} the chain length, \textit{ii)} the overall size of the fluctuation ensemble, \textit{iii)} the accuracy in the search of the equilibrium twist conformation $< h >_{j^{*}}$ i.e., the J-value  \cite{n3}. 
In Section V, the model is applied to two homogeneous fragments with $N=\,11$ and $N=\,41$.
For $N= \,41$, summing over $\sim 10^{8}$ configurations for each dimer in the chain and taking $J=\,60$, the run time is about 22 hours (for each $\Gamma$) on a workstation Intel Xeon E5-1620 v2, 3.7GHz processor.

{While the model here devised can describe in principle any double stranded sequence compatibly with the available computational time, it is recognized that our treatment of the effects of spatial confinement on DNA chains makes only an initial contribution in the field of the mesoscopic Hamiltonian models.
In particular our description applies to diluted crowders distributions whose effects on the DNA conformation can be simulated by a site dependent contraction of the base pair fluctuations as in Eqs.~(\ref{eq:07}),~(\ref{eq:08}).
More refined models could be elaborated in future investigations also in view of the current experimental capability \cite{nakano14,nakano18} to prepare confining networks which simulate the highly crowded conditions of the cellular environment. In this regard one viable strategy  may be that of incorporating the crowders effects in the Hamiltonian model itself through a specific potential term rather than tuning phenomenologically the base pair cutoffs.
}

\section*{V. Results}

\begin{figure}
\includegraphics[height=14.0cm,width=14.0cm,angle=-90]{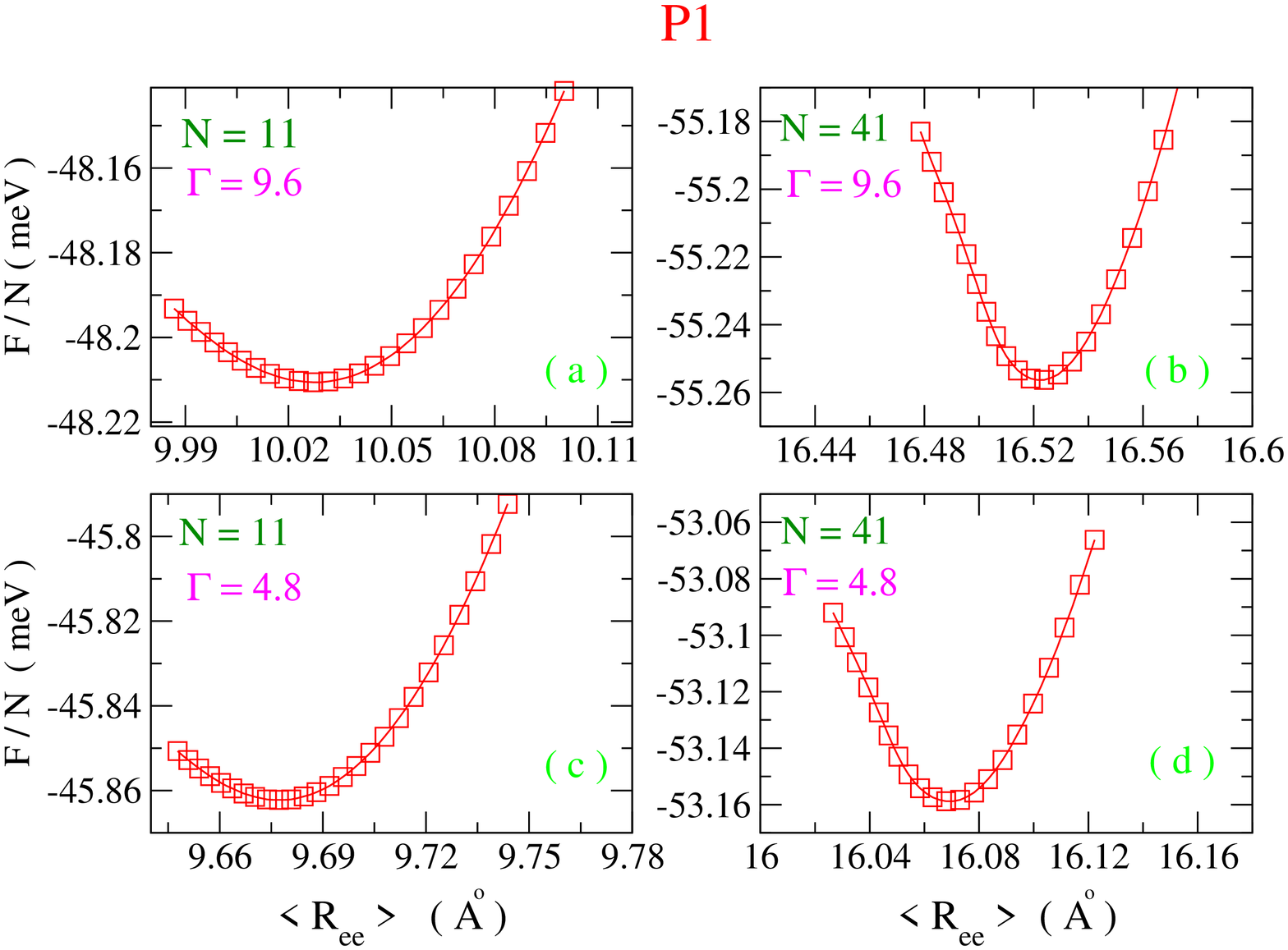}
\caption{\label{fig:3}(Color online)  
Free energy per base pair as a function of the average end-to-end distance, calculated for two short homogeneous fragments at room temperature. The crowders around the fragments are arranged as in Fig.~\ref{fig:2}(a). In (a) and (b),  the mid-chain cutoff on the base pair fluctuation is $\Gamma =\,9.6$. In (c) and (d), the confinement due to the crowders is assumed to be stronger i.e., $\Gamma =\,4.8$. }
\end{figure}

The free energy per base pair is plotted in Fig.~\ref{fig:3} assuming for both molecules the crowders distribution of Fig.~\ref{fig:2}(a). For each molecule two $\Gamma$ values are considered, the smaller value indicating a larger crowders size hence, a stronger confinement for the base pair fluctuations. The calculation is performed at $T=\,300K$. 
For all panels, we report the range of average end-to-end distances around the minimum for $F/N$.  As detailed above, $F/N$ is first computed and minimized as a function of the $< h >_{j}$'s in Eq.~(\ref{eq:09}). For each $< h >_{j}$, the code determines the respective $< R_{e-e} >$ via Eq.~(\ref{eq:10}).
The minimum for  $F/N$ selects the $< R_{e-e} >$ value associated to the twist conformation $< h >_{j^{*}}$ which marks the thermodynamic equilibrium. Note that, by lowering $\Gamma$ for a given chain length (e.g. from panel (a) to (c)), the free energy minimum shifts downwards along the  $< R_{e-e} >$ axis indicating a reduction of the molecule size, at least for these degrees of confinements. Further,  smaller $\Gamma$'s bring about an enhancement of the free energy values (the scales of the y-axis are less negative) consistent with the fact that a larger confinement of the base pair fluctuations drives the system towards a more ordered state. For the spatial ranges reported along the x-axis of all panels, the differences in the free energy scales are small so that thermal fluctuations may bring the molecule from one state to the neighbor one. Note however that: \textit{a)} such differences rapidly grow outside the range visualized in Fig.~\ref{fig:3}; \textit{b)} the energy differences on the ordinate scale should be multiplied by $N$ to evaluate the effect on the whole chain. Then, differences in $F/N$ of $\sim 1meV$ become sizeable and larger than a thermal fluctuation for a chain with e.g., 30 base pairs.

\begin{figure}
\includegraphics[height=14.0cm,width=14.0cm,angle=-90]{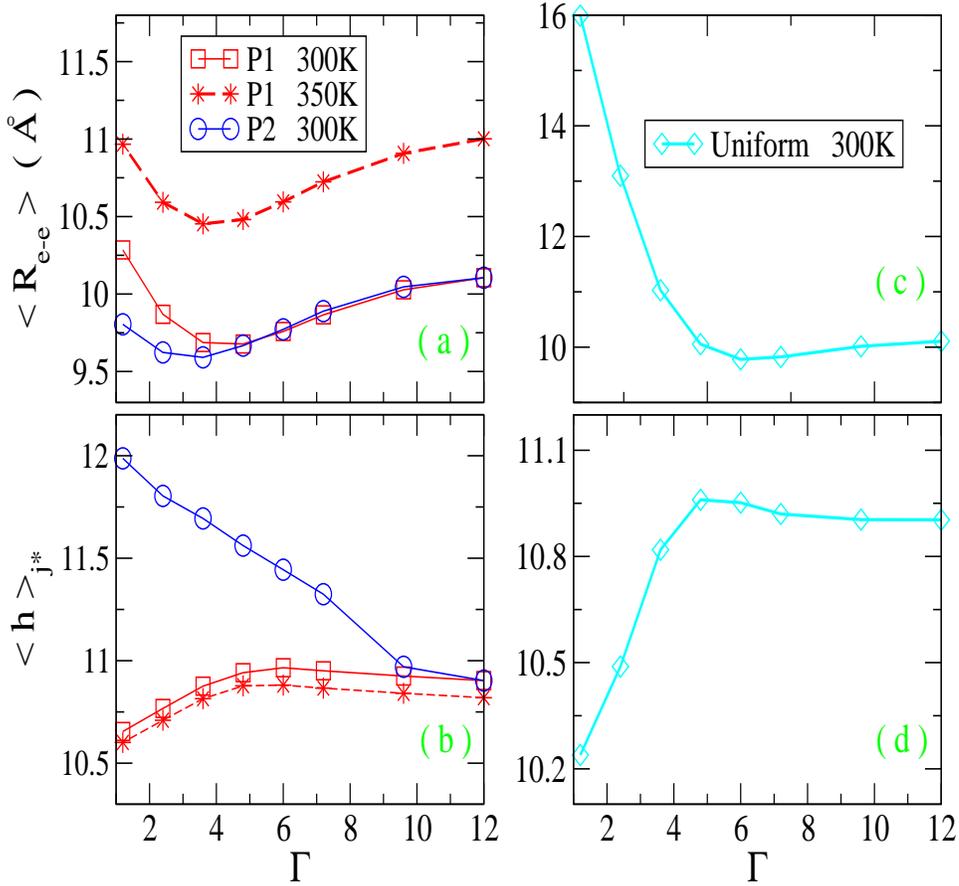}
\caption{\label{fig:4}(Color online)  
 (a), (c) Average end-to-end distance and (b), (d) average equilibrium helical repeat plotted as a function of the cutoff on the base pair fluctuations, for a chain with eleven base pairs.  The values in (a), (c) are calculated  for the equilibrium twist conformations defined respectively in (b), (d).  The simulations, yielding the panels (a) and (b), are carried out at room temperature for both crowders distributions in Fig.~\ref{fig:2} and also at $T=\,350K$ for the profile P1 of Fig.~\ref{fig:2}(a). The panels (c) and (d) are obtained at room temperature by assuming a uniform confinement throughout the chain.  For $\Gamma =\,12$, there is no confinement due to crowders.}
\end{figure}

In Fig.~\ref{fig:4}, we consider the chain with $N=\,11$. The equilibrium twist conformations and the average end-to-end distances are displayed for a set of $\Gamma$'s assuming both the crowders profile with maximum confinement at mid-chain and the profile which mostly confines one chain end. In Fig.~\ref{fig:4}(a), both for P1 and P2, $< R_{e-e} >$ decreases by reducing the cutoff from $\Gamma=\,12$ (zero crowders confinement) to $\Gamma =\,4.8 \,$ i.e., the value considered in Fig.~\ref{fig:3}. However, by further reducing $\Gamma$, $< R_{e-e} >$ turns into a growing function for both profiles although the turning point occurs at a somewhat lower $\Gamma$ value for the P2 plot which also displays a smoother gradient than P1. This interesting feature is physically understood as follows: a moderate degree of confinement shrinks the amplitude of the base pair fluctuations thus reducing the molecule size as measured by $< R_{e-e} >$ but, once the crowders size get larger and the confinement markedly stronger, then the intra-strand base pair distances are stretched and the average end-to-end distance accordingly grows. 

Importantly we also notice that, in the case P1, a high percentage of base pairs around the mid-chain is subjected for small $\Gamma$'s to the strong crowders confinement whereas, in the case P2, only the base pairs in the proximity of one chain end feel the same strong effect. This explains why the distribution P1 stretches the molecule more effectively than P2 for small $\Gamma$'s, the regime where the average end-to-end distance is found to increase. This difference (between the effects caused by P1 and P2) is all the more significant in very short fragments for which a relatively large fraction of base pairs is close to the chain ends.

It is emphasized that the location of the upturn along the $\Gamma$ axis also depends on the assumption that the $C(i)$'s  in Eqs.~(\ref{eq:07}),~(\ref{eq:08}) scale linearly with the site index. Let's take, for instance, a profile with the same chain end conditions as in Eq.~(\ref{eq:08}) but $C(i) \propto \sqrt{i}$. As in this case the crowders would more efficiently shrink the base pair fluctuations throughout the chain, the molecule would more easily elongate than in the case P2. Accordingly the upturn should occur at somewhat larger $\Gamma$ values than in Fig.~\ref{fig:4}(a).

The differences in the molecule behavior due to the two crowders profiles are evident also in Fig.~\ref{fig:4}(b) which plots the average helical repeat at equilibrium. Pinning the molecule mostly at mid-chain (case P1) substantially constrains the helix and prevents significant changes in the helical repeat.  Only when the confinement is sufficiently strong to stretch the chain, as seen in Fig.~\ref{fig:4}(a), the helix is predicted to over-twist. This result hints to a peculiar twist-stretch coupling which in fact has been previously noticed in a different context, that is with an experimental set up which applies a torque in order to over-twist a single molecule in the presence of an external constant force, at least up to $30pN$  \cite{busta06,croq06,bohr11}. 

Instead, pinning the molecule mostly at one chain end (case P2), while letting free the opposite end, has the general effect to untwist the helix. Accordingly, $< h >_{j^{*}}$ grows by decreasing $\Gamma$ through the whole set of values.

One of the advantages of the path integral formalism is that it allows to monitor, through Eq.~(\ref{eq:03}) and the cutoffs in Eq.~(\ref{eq:04}),  the temperature dependence of the macroscopic helical parameters in a crowded environment. Thus, for the crowders distribution P1, we have performed the calculation also at $T=\,350K$ to emphasize the temperature effects with the caveat that in such range short oligomers may be close to the melting also for large salt concentrations \cite{owcz04,pablo07}. 
It is seen in Fig.~\ref{fig:4} that $< R_{e-e} >$ is enhanced with respect to the room temperature value for all $\Gamma$'s consistently with the fact that, at larger $T$, the base pair fluctuations are broader and the stacking bonds are weaker. 
On the other hand, the $< h >_{j^{*}}$ plot lies slightly below the room temperature curve. Again we find that the overall molecule stretching is related to an average helical over-twisting but in this case the two concomitant phenomena are driven by the temperature.

Although the melting transition is not studied here, our results suggest an observation in this regard. Approaching the denaturation temperature, one expects on general grounds that an enhancement of the base pair stretching and stacking fluctuations should cause an average helix untwisting \cite{n2}. Such picture however may be modified by specific crowding conditions (as those fulfilled by P1) that, at increasing $T$, lead to a significant enhancement of the end-to-end distance meanwhile favoring a slight over-twisting of the molecule.

{We also note that the picture associated to the case P1 may provide, in the small $\Gamma$ regime, a reasonable approximation to the nano-channel confinement of a DNA molecule  which causes the observed stretching of the fragments discussed in the Introduction. 
However, as in a nano-channel the molecule is expected to be uniformly confined, we can explicitly simulate such conditions within our model by assuming a constant (i.e., site independent) crowders profile with $C(i)=\,\gamma U$. The results for $< R_{e-e} >$ and $< h >_{j^{*}}$ are displayed in Fig.~\ref{fig:4}(c) and (d), respectively. It is found that, as soon as the confinement becomes sizable ($\Gamma < 5$), the helix markedly over-twists and consistently shows a significant increase in the average end-to-end distance which is stretched to values even larger than $15 \AA$ in the regime in which all base pairs are strongly confined.
}

\begin{figure}
\includegraphics[height=10.0cm,width=10.0cm,angle=-90]{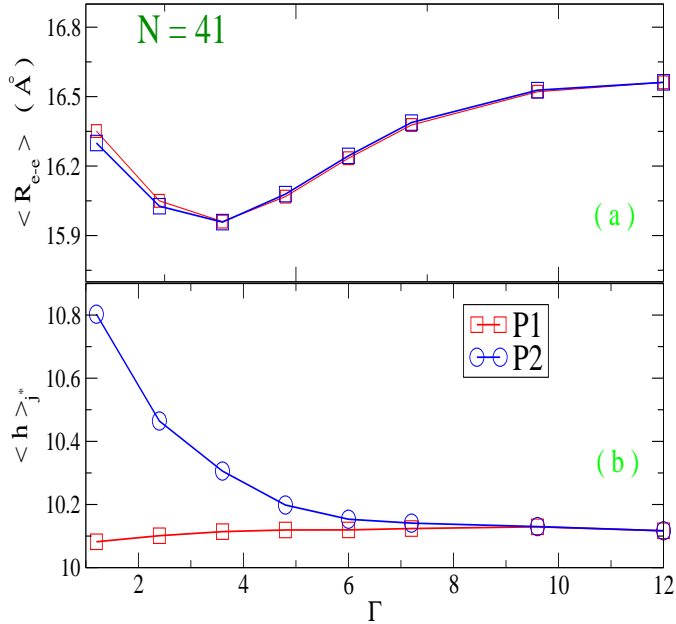}
\caption{\label{fig:5}(Color online)  
 As in Fig.~\ref{fig:4}(a),(b) but for a chain with forty-one base pairs. The calculations are performed at room temperature for both crowders distributions.}
\end{figure}

Finally the computation is performed for a longer chain with $N= \,41$ assuming both crowders profiles. The results, shown in Fig.~\ref{fig:5}, conferm the pattern discussed so far. It is  however significant that the discrepancy in $< R_{e-e} >$ between the two profiles is less pronounced than that noticed for the chain with $N= \,11$ and the upturn occurs now, for both profiles, at the same $\Gamma$. In fact, by increasing the molecule length, the chain end effects should become less relevant. Accordingly, the differences in the confinement caused by the two crowder profiles are expected to be less pronounced, at least with regard to the average end-to-end distance. Instead, as seen in Fig.~\ref{fig:5}(b), the helix untwisting promoted by a sizeable confinement in the case P2  has no analogous in the case P1: for the latter the chain slightly over-twists by increasing the confinement. Thus the different helical behavior is similar to that discussed for the shorter chain.

\section*{VI. Conclusions}

We have studied the behavior of short DNA fragments in equilibrium with an environment populated by objects of variable size. While the crowders generally shrink the real space available to the base pair fluctuations, two particular crowders distribution have been chosen to analyze the molecule response to a confinement which is not uniform throughout the chain. The first crowders profile (P1) assumes that the confinement is maximum at the mid-chain and drops linearly towards the chain ends. The second profile (P2) assumes that the confinement is maximum for the last base pair in the chain and decreases linearly towards the opposite chain end.
The analysis has been carried out by computing two macroscopic parameters providing a complementary, albeit independent, picture of the molecule shape i.e., the average end-to-end distance which measures the fragment size and the average helical repeat which defines the twist conformation. For each crowders distribution,  we have tuned the strength of the confinement and calculated such macroscopic parameters at the thermodynamic equilibrium state, selected among a range of possible helical conformations.

We model the fragments by a three dimensional mesoscopic Hamiltonian incorporating both the radial base pair fluctuations and the angular fluctuations which twist and bend adjacent base pairs along the molecule backbone.
The computation is performed via a path integration method based on the assumption that the base pair distances are temperature dependent trajectories. Summing over a large ensemble of path configurations for the vibrating dimers in the chain, one constructs the average macroscopic parameters. Thus the ensemble averaged helical conformations also include the effects of the base pair radial and bending fluctuations.
The results here obtained follow from a computational technique which accurately performs a site specific confinement of the base pair vibrations.

Applying the method to two short fragments, we have found that a confinement of moderate strength steadily reduces the average end-to-end distance but, once the confinement becomes strong, both chains finally stretch. This effect is more pronounced for the shorter chain whose end-to-end distance (with confinement P1) becomes even larger than in the initial, uncrowded condition. While single molecule experiments on kilo-base long DNA have shown that helical molecule should elongate when over-twisted, our results posit that the interplay between twist and stretching may be more complex in crowded environments. In particular, a strong confinement with largest value at mid-chain (P1) over-twists the  fragments (mostly the shorter one) which indeed extends its end-to-end distance. 
{A similar pattern, with even more pronounced molecule stretching and over-twisting, is found in the case of a uniform confinement throughout the chain, a model appropriate to describe genome fragments in nano-channel arrays.
}
Instead, if the confinement is largest at one chain end (P2) then the molecule elongation is accompanied by a strong helical untwisting and this effect occurs for both fragments.
These findings point to a peculiar correlation between molecule shape and environment which could be directly controlled by engineering specific confining devices.
In this regard, it seems worth to extend the analysis to a variety of crowders distributions and to longer molecules whose behavior in confining environments may be more easily accessed by experiments.


\begin{thebibliography}{widest-label}
\bibitem{n1}
Imposing this condition, the code rejects those short range fluctuations which would bring the pair mates in close proximity, i.e.,  $|r_i| - R_0 < - \ln 2 / b_i$, where  $b_i$ is the Morse potential width.
\bibitem{n3}
More precisely, the code first searches the free energy minimum taking a set of $h$-values in a broad range around $h^{exp}$, say $[8, 12]$, with a partition step $\Delta h_1$. After selecting the first $< h >_{j^{*}}$ for a given $h_{j}$, the code does a second search around $h_{j}$ taking a narrower range and with a finer partition step \, $\Delta h_2 < \Delta h_1$. The procedure is repeated until the equilibrium average helical repeat converges i.e., it becomes independent of the partition step. The  J-value mentioned in the text refers to the last search.
\bibitem{n2}
A ladder Hamiltonian model is a physically meaningful representation only in the melting regime whereby the two complementary strands are moving apart. Instead, in our calculation, the DNA chain fluctuates within a finite range of average twist values tuned by the crowding conditions. Thus the molecule is always assumed to exist in a helical conformation.
\end{thebibliography}
\end{document}